\documentclass[graybox]{svmult}

\usepackage{type1cm}        
\usepackage{makeidx}         
\usepackage{graphicx}        
\usepackage{multicol}        
\usepackage[bottom]{footmisc}

\usepackage{newtxtext}       %
\usepackage{newtxmath}       

\usepackage{epsfig,endnotes}
\usepackage[numbers,sort&compress]{natbib}
\usepackage{multirow}
\usepackage{subfig}
\usepackage{appendix}
\usepackage{graphicx}
\usepackage{grffile}

\newcounter{subcopyrightbox@save}
\usepackage[font=bf]{caption}
\usepackage{color, url}
\usepackage{xspace} 
\usepackage{mathrsfs}
\usepackage{amssymb}
\usepackage{amsmath}
\usepackage{epstopdf}
\usepackage{balance}
\usepackage{algorithm}
\usepackage{algorithmic}

\usepackage{bbold}
\newcommand{\argmax}{\operatornamewithlimits{argmax}}
\newcommand{\argmin}{\operatornamewithlimits{argmin}}
\newcommand{\myparatight}[1]{\smallskip\noindent{\bf {#1}:}~}

\graphicspath{ {../fig/} }

\makeindex 

\begin{document}
%
\title*{Defending against Machine Learning based Inference Attacks via Adversarial Examples: Opportunities and Challenges}
\titlerunning{Defending against Inference Attacks via Adversarial Examples}

\author{
{\rm Jinyuan Jia} 
\and
{\rm  Neil Zhenqiang Gong} 
}
\institute{ Jinyuan Jia \at  Duke University, Durham, NC 27708, \email{jinyuan.jia@duke.edu}
\and Neil Zhenqiang Gong \at Duke University, Durham, NC 27708, \email{neil.gong@duke.edu} }
\maketitle

\abstract{As machine learning (ML) becomes more and more powerful and easily accessible, attackers increasingly leverage ML to perform automated large-scale inference attacks in various domains. In such an ML-equipped inference attack, an attacker has access to some data (called \emph{public data}) of an individual, a software, or a system; and the attacker uses an ML classifier to automatically infer their private data. Inference attacks pose severe privacy and security threats to individuals and systems. Inference attacks are successful because private data are statistically correlated with public data, and ML classifiers can capture such statistical correlations.  In this chapter, we discuss the opportunities and challenges of defending against ML-equipped inference attacks via \emph{adversarial examples}. Our key observation is that attackers rely on ML classifiers in inference attacks. The adversarial machine learning community has demonstrated that ML classifiers have various vulnerabilities. Therefore, we can turn the vulnerabilities of ML into defenses against inference attacks. For example, ML classifiers are vulnerable to adversarial examples, which add carefully crafted noise to normal examples such that an ML classifier makes predictions for the examples as we desire. To defend against inference attacks, we can add carefully crafted noise into the public data to turn them into adversarial examples,  such that  attackers' classifiers make incorrect predictions for the private data. However, existing methods to construct adversarial examples are insufficient because they did not consider the unique challenges and requirements for the crafted noise at defending against inference attacks. In this chapter, we take defending against inference attacks in online social networks as an example to illustrate the opportunities and challenges. 
}

\section{Introduction}



As ML provides more and more powerful tools for data analytics, attackers increasingly use ML to perform automated inference attacks in various domains. Roughly speaking, an ML-equipped inference attack aims to infer private data of an individual, a software, or a system via leveraging ML classifiers to automatically analyze their certain public data. Inference attacks pose pervasive privacy and security threats to individuals and systems. Example inference attacks include, but are not limited to, \emph{attribute inference attacks} in online social networks~\cite{Otterbacher10,weinsberg2012blurme,Zheleva09,Chaabane12,kosinski2013private,Gong14,GongAttriInferSEC16,AttriInfer,Gong18tops,Zhang18}, \emph{author identification attacks}~\cite{narayanan2012feasibility,Payer15,Caliskan15,Caliskan18,shetty2018a4nt,Abuhamad18}, \emph{website fingerprinting attacks} in anonymous communications~\cite{herrmann2009website,panchenko2011website,cai2012touching,juarez2014critical,wang2014effective}, \emph{side-channel attacks} to steal a system's cryptographic keys~\cite{Lerman11,Zhang12}, \emph{membership inference attacks}~\cite{SSSS17,NSH18,SZHBFB19}, \emph{sensor-based location inference attacks}~\cite{MichalevskyPowerSEC,Narain16Oakland}, \emph{feature inference attacks}~\cite{fredrikson2014privacy,Yeom18}, \emph{CAPTCHA breaking attacks}~\cite{ye2018yet,breakAudioCaptchaBurszteinSP,breakTextCaptchaBurszteinCCS11}, etc.. For instance, in  attribute inference attacks, an attacker uses ML classifiers to infer online social network users' private attributes (i.e., private data)--such as location, gender, sexual orientation, and political view--from their publicly available data such as page likes, rating scores, and social friends. The Facebook data privacy scandal in 2018 is a real-world example of attribute inference attack, where Cambridge Analytica leveraged an ML classifier to infer a large amount of Facebook users' private attributes via their public page likes~\cite{Cambridge}.

Existing defenses against inference attacks can be roughly grouped into two categories, i.e., \emph{game-theoretic methods}~\cite{ShokriCCS12,ShokriPETS15,privacygamelocation16,Calmon:2012,AttriGuard} and \emph{(local) differential privacy}~\cite{Dwork:2006,RandomizedResponse,DuchiLDP13,Erlingsson:2014,BassilySuccinctHistograms15,Wang17LDP,Calibrate}.  
These methods are not practical: they  either are computationally intractable or incur large utility loss of the public data. 
Specifically, in game-theoretic methods, an attacker performs the optimal inference attack based on the knowledge of the defense, while the defender defends against the optimal inference attack. These methods are computationally intractable. In particular, the computation cost to find the noise that should be added to the public data is \emph{exponential} to the dimensionality of the public data~\cite{AttriGuard}, which is often large in practice. Salamatian et al.~\cite{Salamatian:2015} proposed to quantize the public data to approximately solve the game-theoretic optimization problem~\cite{Calmon:2012}. However, quantization incurs a large utility loss of the public data~\cite{AttriGuard}. Differential privacy (DP)~\cite{Dwork:2006} or its variant called local differential privacy (LDP)~\cite{RandomizedResponse,DuchiLDP13,Erlingsson:2014,BassilySuccinctHistograms15,Wang17LDP,Calibrate} can also be applied to defend against inference attacks. DP/LDP provides a strong privacy guarantee. However, DP/LDP aims to achieve a privacy goal that is different from the one in inference attacks.  For instance, 
LDP's privacy goal is to add random noise to a true public data record such that two arbitrary true public data records have close probabilities (their difference is bounded by a privacy budget)   
to generate the same noisy public data record. However, in defending against inference attacks, the privacy goal is to add noise to a public data record 
such that the private data cannot be accurately inferred by the attacker's classifier. 
As a result, DP/LDP achieves a suboptimal privacy-utility tradeoff at defending against inference attacks~\cite{AttriGuard}.

In this chapter, we discuss the opportunities and challenges of defending against inference attacks via adversarial examples. Our key intuition is that attackers use ML classifiers in inference attacks. ML classifiers were shown to be vulnerable to adversarial examples~\cite{barreno2006can,Biggio:ECMLPKDD:13,Goodfellow:ICLR:14b}, which add carefully crafted noise to normal examples such that an ML classifier makes predictions for the examples as we desire. Based on this observation, we can add carefully crafted noise to a public data record to turn it into an adversarial example such that attacker's classifiers incorrectly predict the private data. However, existing methods to construct adversarial examples are insufficient because they did not consider the unique challenges of defending against inference attacks. Specifically, the first challenge is that the defender does not know the the attacker's classifier, since there are many possible choices for the classifier. Second, in certain application domains, the true private data value is not available when  turning the public data into an adversarial example, i.e., the ``true label''  is not known when constructing the adversarial example. Third, the noise  added to the public data should satisfy certain utility-loss constraints, which may be different for inference attacks in different application domains. 
We take AttriGuard~\cite{AttriGuard},\footnote{Code and dataset of AttriGuard are publicly available: https://github.com/jjy1994/AttriGuard} the first adversarial example based practical defense against  inference attacks (in particular, attribute inference attacks in online social networks), as an example to illustrate how to address the challenges. 

AttriGuard works in two phases.  Phase I addresses the first two challenges, while Phase II addresses the third challenge. Specifically, in Phase I,  the defender  itself learns an ML classifier to perform attribute inference. The defender constructs adversarial examples based on its own classifier and the adversarial examples are also likely to be effective for the attacker's classifier because of \emph{transferability}~\cite{Goodfellow:ICLR:14b,PracticalBlackBox17,liu2016delving,CarliniSP17}. For each possible attribute value, the defender finds a carefully crafted noise vector to turn a user's public data into an adversarial sample, for which the defender's classifier predicts the attribute value.  Existing adversarial example methods (e.g.,~\cite{Goodfellow:ICLR:14b,Papernot:arxiv:16Limitation,sharif2016accessorize,liu2016delving,CarliniSP17}) are insufficient because they did not consider users' privacy preference. In particular,  different users may have different preferences on what types of noise can be added to their public data. For instance, a user may prefer modifying its existing rating scores, or adding new rating scores, or combination of them. To address the challenge, AttriGuard optimizes the method developed by Papernot et al.~\cite{Papernot:arxiv:16Limitation} to incorporate such constraints.

In Phase II, the defender randomly picks a noise vector found in Phase I according to a probability distribution $\mathbf{q}$ over the noise vectors. AttriGuard finds the probability distribution $\mathbf{q}$ via minimizing its distance to a \emph{target probability distribution $\mathbf{p}$} subject to a bounded utility loss of the public data. The target probability distribution is selected by the defender. For instance, the target probability distribution could be a uniform distribution over the possible attribute values, with which the defender aims to make the attacker's inference close to random guessing. Formally, AttriGuard formulates finding the probability distribution $\mathbf{q}$ as solving a constrained convex optimization problem. Moreover, according to the \emph{Karush-Kuhn-Tucker (KKT) conditions}~\cite{convexOptimization}, the solution of the probability distribution $\mathbf{q}$ is intuitive and interpretable.

Jia and Gong~\cite{AttriGuard} compared AttriGuard with existing defenses using a real-world dataset~\cite{Gong12-imc,GongAttriInferSEC16}. In the dataset, a user's public data are the rating scores the user gave to mobile apps on Google Play, while the attribute is the city a user lives/lived in. First, the results demonstrated that the adapted adversarial example method in Phase I outperforms existing ones. Second, AttriGuard is effective at defending against attribute inference attacks. For instance, by modifying at most 4 rating scores on average, the attacker's inference accuracy is reduced by 75\% for several defense-unaware attribute inference attacks and attacks that adapt to the defense. 
Third, AttriGuard adds significantly smaller noise to users' public data than existing defenses when reducing the attacker's inference accuracy by the same amount. 

In the rest of this chapter, we will review inference attacks and their defenses, how to formulate the problem of defending against inference attacks, and describe AttriGuard to solve the formulated problem. Finally, we will discuss further opportunities, challenges, and future work for leveraging adversarial machine learning to defend against inference attacks. 

%
%
%

\section{Related Work}
\label{relatedwork}

\subsection{Inference Attacks}
\label{sec:attacks}

\begin{table}[!t]\renewcommand{\arraystretch}{1}
\centering
\caption{Some inference attacks.}
\begin{tabular}{|c|c|c|} \hline 
Attack &  Public data & Private data \\ \hline
\multirow{3}{*}{Attribute inference attacks~\cite{Otterbacher10,weinsberg2012blurme,Zheleva09,Chaabane12,kosinski2013private,Gong14,GongAttriInferSEC16,AttriInfer,Gong18tops,Zhang18}} &  Rating scores,  & Age,  \\ 
&   page likes,  &  gender,  \\ 
&   social friends, etc. &  political view, etc. \\ \hline
Author identification attacks~\cite{narayanan2012feasibility,Payer15,Caliskan15,Caliskan18,shetty2018a4nt,Abuhamad18} &  Text document, program & Author identity \\ \hline
Website fingerprinting attacks~\cite{herrmann2009website,panchenko2011website,cai2012touching,juarez2014critical,wang2014effective} &  Network traffic & Website \\ \hline
\multirow{3}{*}{Side-channel attacks~\cite{Lerman11,Zhang12}} &  Power consumption,  & \multirow{3}{*}{Cryptographic keys} \\ 
 &   processing time,  &  \\ 
 &   access pattern &  \\ \hline
Membership inference attacks~\cite{SSSS17,NSH18,SZHBFB19} &  Confidence scores, gradients & Member/Non-member \\ \hline
Location inference attacks~\cite{MichalevskyPowerSEC,Narain16Oakland} &  Sensor data on smartphone & Location \\ \hline
Feature inference attacks~\cite{fredrikson2014privacy,Yeom18} &  Partial features, model prediction & Missing features \\ \hline
CAPTCHA breaking attacks~\cite{ye2018yet,breakAudioCaptchaBurszteinSP,breakTextCaptchaBurszteinCCS11} &  CAPTCHA & Text, audio, etc. \\ \hline
\end{tabular} 
\label{attacks} 
\end{table}

\myparatight{Attribute inference attacks}
A number of  studies~\cite{Otterbacher10,weinsberg2012blurme,Zheleva09,Chaabane12,kosinski2013private,Gong14,GongAttriInferSEC16,AttriInfer,Gong18tops,Zhang18} have demonstrated that users in online social networks are vulnerable to \emph{attribute inference attacks}. 
In these attacks, an attacker has access to a set of  data (e.g., rating scores, page likes, social friends) about a target user, which we call \emph{public data}; 
and the attacker aims to infer \emph{private attributes} (e.g., location, political view, or sexual orientation)
of the target user. Specifically, the attacker could be a social network provider, advertiser, or data broker (e.g., Cambridge Analytica). 
The attacker first collects a dataset from users who disclose both their public data and attributes, and then the attacker uses them as a training dataset to learn a ML classifier, which takes a user's public data as input and predicts the user's private attribute value. Finally, the attacker applies the classifier to predict the target user's private attribute value via its public data. 
Attribute inference attacks can accurately infer users' various private attributes via their publicly available rating scores, page likes, and social friends. For instance, Cambridge Analytica  inferred Facebook users' various private attributes including, but not limited to, age, gender, political view, religious view, and relationship status, using their public page likes~\cite{Cambridge}.

\myparatight{Other inference attacks} Other than attribute inference attacks in online social networks, inference attacks are also pervasive in other domains. Table~\ref{attacks} shows some inference attacks and the corresponding public data and private data.  In author identification attacks~\cite{narayanan2012feasibility,Payer15,Caliskan15,Caliskan18,shetty2018a4nt,Abuhamad18}, an attacker can identify the author(s)  of an anonymous text document or program via leveraging ML to analyze the writing style. In website fingerprinting attacks~\cite{herrmann2009website,panchenko2011website,cai2012touching,juarez2014critical,wang2014effective}, an attacker can infer the website a user visits via leveraging ML to analyze the network traffic, even if the traffic is encrypted and anonymized. In side-channel attacks~\cite{Lerman11,Zhang12}, an attacker can infer a system's cryptographic keys via leveraging ML to analyze the power consumption, processing time, and access patterns. In membership inference attacks~\cite{SSSS17,NSH18,SZHBFB19}, an attacker can infer whether a data record is in a classifier's training dataset via leveraging ML to analyze the confidence scores of the data record predicted by the classifier or the gradient of the classifier with respect to the data record. In sensor-based location inference attacks~\cite{MichalevskyPowerSEC,Narain16Oakland}, an attacker can  infer a user's locations via leveraging ML to analyze  the user's  smartphone's aggregate power consumption as well as the gyroscope, accelerometer, and magnetometer
data available from the user's smartphone. In feature inference attacks~\cite{fredrikson2014privacy,Yeom18},\footnote{These attacks are also called attribute inference attacks~\cite{Yeom18}. To distinguish with  attribute inference attacks in online social networks, we call them feature inference attacks.} an attacker can infer a data point's missing features (e.g., an individual's genotype) via analyzing an ML model's prediction for the data point. In CAPTCHA breaking attacks~\cite{ye2018yet,breakAudioCaptchaBurszteinSP,breakTextCaptchaBurszteinCCS11}, an attacker can solve a CAPTCHA via ML. 

\subsection{Defenses}

\myparatight{Game-theoretic methods and differential privacy}
Shokri et al.~\cite{ShokriCCS12}, Calmon et al.~\cite{Calmon:2012}, and Jia and Gong~\cite{AttriGuard} proposed  game-theoretic methods to defend against  inference attacks. These methods  rely on optimization problems that are computationally intractable when the public data is high dimensional. 
Salamatian et al.~\cite{Salamatian:2015} proposed \emph{Quantization Probabilistic Mapping (QPM)} to approximately solve the game-theoretic optimization problem formulated by Calmon et al.~\cite{Calmon:2012}. Specifically, they cluster public data and use the cluster centroids to represent them. Then, they approximately solve the optimization problem using the cluster centroids. Huang et al.~\cite{GAP} proposed to use generative adversarial networks to approximately solve the game-theoretic optimization problems. However, these approximate solutions do not have formal guarantees on utility loss of the public data. Differential privacy or local differential privacy~\cite{Dwork:2006,RandomizedResponse,DuchiLDP13,Erlingsson:2014,BassilySuccinctHistograms15,Wang17LDP,Calibrate} can also be applied to add noise to the public data to defend against inference attacks. However, as we discussed in Introduction, they achieve suboptimal privacy-utility tradeoffs because they aim to provide privacy guarantees that are stronger than needed to defend against inference attacks.

\myparatight{Other methods} 
Other methods~\cite{weinsberg2012blurme,ChenObfuscationPETS14}
 leveraged heuristic correlations between the entries of the public data and attribute values to defend against attribute inference attacks in online social networks. Specifically, they modify the $k$ entries that have large correlations with the attribute values that do not belong to the target user. $k$ is a parameter to control privacy-utility tradeoffs. 
For instance, Weinsberg et al.~\cite{weinsberg2012blurme} proposed BlurMe, which calculates the correlations based on the coefficients of a logistic regression classifier that models the relationship between public data entries and attribute values.   
 Chen et al.~\cite{ChenObfuscationPETS14} proposed ChiSquare, which computes the correlations between public data entries and attribute values based on chi-square statistics. 
These methods suffer from two limitations: 1) they require the defender to have direct access to users' private attribute values, which makes the defender become a single point of failure, i.e., when the defender is compromised, the private attribute values of all users are compromised; and 2)  they incur large utility loss of the public data.

\section{Problem Formulation}
\label{problem}

%


We take attribute inference attacks in online social networks as an example to illustrate how to formulate the problem of defending against inference attacks. However, our problem formulation can also be generalized to other inference attacks. We have three parties: \emph{user}, \emph{attacker}, and \emph{defender}.  
Next, we discuss each party one by one.

\subsection{User} 
We focus on  protecting the private attribute of one user. 
We can protect different users separately. 
A user aims to publish some data while preventing inference of its private attribute from the public data. 
We denote the user's public data and private attribute as $\mathbf{x}$ (a column vector) and $s$, respectively. 
 For instance,  an entry of the public data vector $\mathbf{x}$ could be the rating score the user gave to an item or 0 if the user did not rate the item; an entry of the public data vector could also be 1 if the user liked the corresponding page or 0 otherwise. 
 For simplicity, we assume each entry of $\mathbf{x}$ is normalized to be in the range $[0,1]$.
The  attribute $s$ has $m$ possible values, which we denote as $\{1,2,\cdots,m\}$; $s=i$ means that the user's private attribute value is $i$.
For instance, when the private attribute is political view, the attribute could have two possible values, i.e., democratic and republican. 
We note that the attribute $s$ could be a combination of multiple attributes. 
For instance, the attribute could be $s=(\text{political view, gender})$, which has four possible values, i.e., (democratic, male), (republican, male), (democratic, female), and (republican, female).

\myparatight{Policy to add noise} Different users may have different preferences over what kind of noise can be added to their public data.  For instance, a user may prefer modifying its existing rating scores, 
while another user may prefer adding new rating scores.
We call a policy specifying what kind of noise can be added a \emph{noise-type-policy}. 
In particular, we consider the following three types of  noise-type-policy.
\begin{itemize}
\item {\bf Policy A: Modify\_Exist.} In this policy, the defender can only modify the non-zero entries of $\mathbf{x}$. When the public data are rating scores, this policy  means that the defender can only modify a user's existing rating scores. When the public data correspond to page likes, this policy means that  the defender can only remove a user's existing page likes. 
\item {\bf Policy B: Add\_New.} In this policy, the defender can only change the zero entries of $\mathbf{x}$. When the public data are rating scores, this policy means that the defender can only add new rating scores for a user. When the public data represent page likes, this policy means that  the defender can only add new page likes for a user. We call this policy \emph{Add\_New}.
\item {\bf Policy C: Modify\_Add.} This policy is a combination of Modify\_Exist and Add\_New. In particular, the defender could modify any entry of $\mathbf{x}$. 
\end{itemize}

\subsection{Attacker} 

The attacker has access to the noisy public data and aims to infer the user's private attribute value. 
We consider an attacker has 
 an {ML classifier} that takes a user's (noisy) public data as input and infers the user's private attribute value. 
Different users might treat different attributes as private. In particular, some users do not treat the attribute $s$ as private, so they publicly disclose it. 
Via collecting data from such users, the attacker can learn the ML classifier. 

We denote the attacker's  classifier as $C_a$, and $C_a(\mathbf{x})\in$ $\{1,2,\cdots,m\}$ is the predicted attribute value for the user whose public data is $\mathbf{x}$. 
The attacker could use a standard ML classifier, e.g., logistic regression, random forest, and neural network. 
Moreover, an attacker can also adapt its attack based on the defense. For instance, the attacker could first try detecting the noise and then perform attribute inference attacks. We assume the attacker's classifier  is unknown to the defender, since there are many possible choices for the attacker's classifier.

\subsection{Defender} 
The defender adds noise to a user's true public data according to a noise-type-policy. The defender can be a software on the user's client side, e.g., a browser extension.
The defender has access to the user's true public data $\mathbf{x}$. The defender adds a  noise vector $\mathbf{r}$ to $\mathbf{x}$ and the attacker has access to the noisy public data vector $\mathbf{x+r}$. The defender aims to achieve two goals:
\begin{itemize}
\item {\bf Goal I.} The attacker's  classifier is inaccurate at inferring the user's private attribute value.
\item {\bf Goal II.} The utility loss of the public data vector is bounded. 
\end{itemize}
However, achieving the two goals faces several challenges. 

\myparatight{Achieving Goal I} The first challenge to achieve Goal I is that the defender does not know the attacker's classifier.  
To address the challenge, the defender itself trains a classifier $C$ to perform attribute inference using the data from the users who share both public data and attribute values. Then, the defender adds a noise vector to a user's public data vector such that its own classifier is inaccurate at inferring the user's private attribute value. We consider the defender's classifier $C$ is implemented in the popular \emph{one-vs-all} paradigm. Specifically, the classifier has $m$ decision functions denoted as $C_1$, $C_2$, $\cdots$, $C_m$, where $C_i(\mathbf{x})$ is the confidence that the user has an attribute value $i$. The classifier's inferred attribute value is  $C(\mathbf{x})=\argmax_i C_i(\mathbf{x})$. Note that, when the attribute only has two possible values (i.e., $m=2$), we have $C_2(\mathbf{x})=-C_1(\mathbf{x})$ for classifiers like logistic regression and SVM.

The second challenge is that the defender does not know the user's true private attribute value. Therefore, for a given noise vector, the defender does not know whether its own classifier makes incorrect prediction on the user's private attribute value or not. One method to address the challenge is to  add noise vectors to users' public data vectors such that the defender's classifier always predicts a certain attribute value for the users. However, for some users, such method may need noise that violates the utility-loss constraints of the public data. 
Therefore,  the defender adopts a \emph{randomized noise addition mechanism} denoted as $\mathcal{M}$. Specifically, given a user's true public data vector $\mathbf{x}$, the defender samples a noise vector  $\mathbf{r}$ from the space of possible noise vectors with a probability $\mathcal{M}(\mathbf{r}|\mathbf{x})$ and adds it to the true public data vector.  

Since the defender adds random noise, the defender's classifier's prediction is also random. We denote by $\mathbf{q}$ the probability distribution of the attribute values predicted by the defender's classifier for the particular user when the defender adds random noise to the user's public data vector according to a randomized noise addition mechanism $\mathcal{M}$. Moreover, the defender aims to find a mechanism $\mathcal{M}$ such that the output probability distribution $\mathbf{q}$ is the closest to a \emph{target probability distribution $\mathbf{p}$} subject to a utility-loss budget, where $\mathbf{p}$ is selected by the defender. For instance, without knowing anything about the attributes, the target probability distribution could be the uniform distribution over the $m$ attribute values, with which the defender aims to make the attacker's inference close to random guessing. The target probability distribution could also be estimated from the users who publicly disclose the attribute, e.g., the probability $p_i$ is the fraction of such users who have attribute value $i$. Such target probability distribution naturally represents a baseline attribute inference attack. The defender aims to reduce an attack to the baseline attack with such target probability distribution. 

The next challenge is how to  quantify the distance between  $\mathbf{p}$ and $\mathbf{q}$. While any distance metric could be applied, AttriGuard  measures the distance between $\mathbf{p}$ and $\mathbf{q}$ using their Kullback--Leibler (KL) divergence, i.e., $KL(\mathbf{p}||\mathbf{q})$=$\sum_i p_i\text{log}\frac{p_i}{q_i}$. AttriGuard chooses KL divergence because it makes the formulated optimization problem convex, which has efficient, accurate, and interpretable solutions.

\myparatight{Achieving Goal II} The key challenge to achieve Goal II is how to quantify the utility loss of the public data vector. 
A user's (noisy) public data are often leveraged by a service provider to provide services. For instance,  
when the public data are rating scores, they are often used to recommend items to users that match their personalized preferences. Therefore, utility loss of the public data can essentially be measured by the service quality loss. 
Specifically, 
 in a recommender system, the decreased accuracy of the recommendations introduced by the added noise can be used as utility loss. However, using such service-dependent utility loss makes the formulated optimization problem  computationally intractable.   

 Therefore, we resort to  utility-loss metrics that make the formulated optimization problems tractable but can still well approximate the utility loss for different services. 
In particular, we can use a distance metric $d(\mathbf{x}, \mathbf{x} + \mathbf{r})$ to measure utility loss. Since $\mathbf{r}$ is a random value generated according to the mechanism $\mathcal{M}$, we can measure the utility loss using the expected distance $E(d(\mathbf{x}, \mathbf{x} + \mathbf{r}))$.  For instance, the distance metric can be $L_0$ norm of the noise, i.e., $d(\mathbf{x}, \mathbf{x} + \mathbf{r})=||\mathbf{r}||_0$. 
$L_0$ norm is the number of entries of $\mathbf{x}$ that are modified by the noise, which has semantic interpretations in a number of real-world application domains. 
For instance, when the public data are rating scores, $L_0$ norm means the number of items whose rating scores are modified. Likewise, when the public data are page likes, an entry of $\mathbf{x}$ is 1 if the user liked the corresponding page, otherwise the entry is 0. Then, $L_0$ norm means the number of page likes that are removed or added by the defender. 
The distance metric can also be $L_2$ norm of the noise, which considers the magnitude of the modified rating scores when the public data are rating scores.

\myparatight{Attribute-inference-attack defense problem} With quantifiable defender's goals, we can formally define the problem of defending against attribute inference attacks. Specifically, the user specifies a noise-type-policy and an utility-loss budget $\beta$. The defender specifies a target probability distribution $\mathbf{p}$, learns a classifier $C$, and finds a mechanism $\mathcal{M}^{*}$, which adds noise to the user's public data such that the user's utility loss is within the budget while the output probability distribution $\mathbf{q}$ of the classifier $C$ is closest to the target probability distribution $\mathbf{p}$. 
 Formally, we have:
\begin{definition}
Given a noise-type-policy $\mathcal{P}$, an utility-loss budget $\beta$, a target probability distribution $\mathbf{p}$, and a classifier $C$, the defender aims to find a mechanism $\mathcal{M}^{*}$ via solving the following optimization problem: 
\begin{eqnarray}
\mathcal{M}^{*}=&\argmin_{\mathcal{M}} KL(\mathbf{p}||\mathbf{q}) \nonumber \\
\label{ulb}
\text{subject to }  & E(d(\mathbf{x}, \mathbf{x} + \mathbf{r})) \leq \beta,
\end{eqnarray}
where $\mathbf{q}$ depends on the classifier $C$ and the mechanism $\mathcal{M}$. AttriGuard uses the $L_0$ norm of the noise as the metric $d(\mathbf{x}, \mathbf{x} + \mathbf{r})$ because of its semantic interpretation. 

\end{definition}

\section{Design of AttriGuard}

\subsection{Overview}
The major challenge to solve the optimization problem in Equation~\ref{ulb} is that the number of parameters of the mechanism $\mathcal{M}$, which maps a given vector to another vector probabilistically, is exponential to the dimensionality of the public data vector.  
To address the challenge, 
Jia and Gong~\cite{AttriGuard} proposed AttriGuard, a \emph{two-phase framework} to solve the optimization problem approximately. 
The intuition is that, although the noise space is large, we can categorize them into $m$ groups depending on the defender's classifier's output. Specifically, 
we denote by $G_i$ 
 the group of noise vectors such that if we add any of them to the user's public data, then the defender's classifier will infer the attribute value $i$ for the user. Essentially, the probability ${q}_i$ that the defender's classifier infers attribute value $i$ for the user is the probability that $\mathcal{M}$ will sample a noise vector in the group $G_i$, i.e., ${q}_i=\sum_{\mathbf{r}\in G_i} \mathcal{M}(\mathbf{r}|\mathbf{x})$. AttriGuard finds one representative noise vector in each group and assumes $\mathcal{M}$ is a probability distribution concentrated on the representative noise vectors.

Specifically, in Phase I, for each group $G_i$, AttriGuard finds a minimum noise $\mathbf{r}_i$ such that if we add $\mathbf{r}_i$ to the user's public data, then the defender's classifier predicts the attribute value $i$ for the user. AttriGuard finds a minimum noise in order to minimize utility loss. 
In \emph{adversarial machine learning}, this is known as \emph{adversarial example}. However, existing adversarial example methods~\cite{Goodfellow:ICLR:14b,Papernot:arxiv:16Limitation,sharif2016accessorize,liu2016delving,CarliniSP17} are insufficient to find the noise vector $\mathbf{r}_i$, because they do not consider the noise-type-policy. AttriGuard optimizes the adversarial example method developed by Papernot et al.~\cite{Papernot:arxiv:16Limitation} to incorporate noise-type-policy. 
The noise $\mathbf{r}_i$ optimized to evade the defender's classifier is also  likely to make the attacker's classifier predict the attribute value $i$ for the user, which is known as \emph{transferability}~\cite{Goodfellow:ICLR:14b,PracticalBlackBox17,liu2016delving,CarliniSP17} in adversarial machine learning. 

In Phase II, AttriGuard simplifies the mechanism $\mathcal{M^*}$ to be a probability distribution over the $m$ representative noise vectors $\{\mathbf{r}_1, \mathbf{r}_2, \cdots, \mathbf{r}_m\}$. In other words, the defender randomly samples a noise vector $\mathbf{r}_i$ according to the probability distribution $\mathcal{M^*}$ and adds the noise vector to the user's public data. Under such simplification, $\mathcal{M^*}$ only has at most $m$ non-zero parameters, the output probability distribution $\mathbf{q}$ of the defender's classifier essentially becomes $\mathcal{M^*}$, and we can transform the optimization problem in Equation~\ref{ulb} to be a convex problem, which can be solved efficiently and accurately. 
 Moreover, Jia and Gong derived the analytical forms of the solution using the \emph{Karush-Kuhn-Tucker (KKT) conditions}~\cite{convexOptimization}, which shows that the solution is intuitive and interpretable. 

\subsection{Phase I: Finding $\mathbf{r}_i$}

 Phase I aims to find a minimum noise $\mathbf{r}_i$ according to the noise-type-policy $\mathcal{P}$, 
such that the classifier $C$ infers the attribute value $i$ for the user after adding $\mathbf{r}_i$ to its public data $\mathbf{x}$. Formally, AttriGuard  finds such  $\mathbf{r}_i$ via solving the following optimization problem:
\begin{eqnarray}
& \mathbf{r}_i = \argmin_{\mathbf{r}} ||\mathbf{r}||_0 \nonumber \\
\label{findr}
\text{subject to } & C(\mathbf{x+r})=i.
\end{eqnarray}

The formulation of finding $\mathbf{r}_i$ can be viewed as finding an {adversarial example}~\cite{barreno2006can,Biggio:ECMLPKDD:13,Goodfellow:ICLR:14b} to the classifier $C$. 
However, existing adversarial example methods (e.g.,~\cite{Goodfellow:ICLR:14b,Papernot:arxiv:16Limitation,sharif2016accessorize,liu2016delving,CarliniSP17}) are insufficient to solve $\mathbf{r}_i$. The key reason is that they do not consider the noise-type-policy, which specifies the types of noise that can be added. 
Papernot et al.~\cite{Papernot:arxiv:16Limitation} proposed a \emph{Jacobian-based Saliency Map  Attack} (JSMA) to deep neural networks.  They demonstrated that JSMA can find small noise (measured by $L_0$ norm) to evade a deep neural network.  Their algorithm iteratively adds noise to an example ($\mathbf{x}$ in our case) until the classifier $C$ predicts $i$ as its label or the maximum number of iterations is reached. 
In each iteration, the algorithm picks  one or two entries of $\mathbf{x}$ based on saliency map, and then increase or decrease the entries by a {constant} value. 

Jia and Gong also designed their algorithm, which is called  \emph{\underline{P}olicy-\underline{A}ware \underline{N}oise Fin\underline{d}ing \underline{A}lgorithm (PANDA)}, based on saliency map. However, PANDA is different from JSMA in two aspects.  First, PANDA incorporates the noise-type-policy, while JSMA does not. The major reason is that JSMA is not developed for preserving privacy, so JSMA does not have noise-type-policy as an input. 
Second, in JSMA, all the modified entries of  $\mathbf{x}$ are either increased or decreased. In PANDA, some entries can be increased while other entries can be decreased. 
Jia and Gong demonstrated that PANDA can find smaller noise than JSMA.


\subsection{Phase II: Finding $\mathcal{M}^*$}
After the defender solves $\{\mathbf{r}_1, \mathbf{r}_2, \cdots, \mathbf{r}_m\}$, the defender randomly samples one of them with a certain probability and adds it to the user's public data $\mathbf{x}$.
Therefore, the randomized noise addition mechanism $\mathcal{M}$ is a probability distribution over $\{\mathbf{r}_1, \mathbf{r}_2, \cdots, \mathbf{r}_m\}$, where $\mathcal{M}_i$ is the probability that the defender adds $\mathbf{r}_i$ to $\mathbf{x}$. 
Since $q_i=\text{Pr}(C(\mathbf{x+r})=i)$ and $C(\mathbf{x+r}_i)=i$, we have $q_i=\mathcal{M}_i$, where $i\in \{1,2,\cdots,m\}$. 
Therefore,   the optimization problem in Equation~\ref{ulb} can be transformed to the following optimization problem: 
\begin{eqnarray}
\mathcal{M}^{*}=&\argmin_{\mathcal{M}} KL(\mathbf{p}||\mathcal{M}) \nonumber \\
\label{ulb-1}
\text{subject to }  & \sum_{i=1}^m \mathcal{M}_i ||\mathbf{r}_i||_0 \leq \beta \nonumber \\
& \mathcal{M}_i > 0, \forall i\in \{1,2,\cdots,m\} \nonumber \\
& \sum_{i=1}^m \mathcal{M}_i = 1,
\end{eqnarray}
where AttriGuard uses the $L_0$ norm of the noise as the utility-loss metric $d(\mathbf{x}, \mathbf{x} + \mathbf{r})$ in Equation~\ref{ulb}.  The above optimization problem is convex because its objective function and constraints are convex, which implies that $\mathcal{M}^*$ is a global minimum. Moreover,  many methods can be applied to solve the optimization problem exactly and efficiently. For instance, Jia and Gong~\cite{AttriGuard} described an KKT conditions based method to solve the optimization problem. We can also use the cvxpy package~\cite{cvxpy} to solve the problem. 

\myparatight{Interpreting the mechanism $\mathcal{M}^*$}  
According to the standard  KKT conditions, the solved mechanism satisfies the following equations:
\begin{eqnarray}
\label{kkt11}
&\triangledown_{\mathcal{M}}(KL(\mathbf{p}||\mathcal{M}^{*}) + \mu_0 (\sum_{i=1}^m \mathcal{M}_i^{*} ||\mathbf{r}_i||_0 - \beta) - \sum_{i=1}^m \mu_{i}\mathcal{M}_i^{*} \nonumber \\
& + \lambda (\sum_{i=1}^m \mathcal{M}_i^{*} -1))=0  \\
\label{kkt12}
&\mu_{i}\mathcal{M}_i^{*} = 0, \forall i\in \{1,2,\cdots,m\}  \\
\label{kkt13}
&\mu_0 (\sum_{i=1}^m \mathcal{M}_i^{*} ||\mathbf{r}_i||_0 - \beta) =0,
\end{eqnarray}
where $\triangledown$ indicates gradient, while $\mu_i$ and $\lambda$ are KKT multipliers.  According to Equation~\ref{kkt12} and $\mathcal{M}_i^*>0$, we have $\mu_i = 0, \forall i\in \{1,2,\cdots,m\}$. Therefore, according to Equation~\ref{kkt11}, Jia and Gong derived the following analytical form of the solved mechanism:
\begin{eqnarray}
\label{ex1}
\mathcal{M}_{i}^{*}=\frac{{p}_{i}}{\mu_0 ||\mathbf{r}_i||_0 + \lambda }
\end{eqnarray}

If we do not have the utility-loss constraint $\sum_{i=1}^m \mathcal{M}_i ||\mathbf{r}_i||_0 \leq \beta$ in the optimization problem in Equation~\ref{ulb-1}, then the mechanism $\mathcal{M}^*=\mathbf{p}$ reaches the minimum KL divergence $KL(\mathbf{p}||\mathcal{M})$, where $\mathbf{p}$ is the target probability distribution selected by the defender. In other words, if we do not consider utility loss, the defender samples the noise $\mathbf{r}_i$ with the target probability $p_i$ and adds it to the user's public data. However, when we consider the utility-loss budget, the relationship between the mechanism $\mathcal{M}^*$ and the target probability distribution $\mathbf{p}$ is represented in Equation~\ref{ex1}. In other words, the defender samples the noise $\mathbf{r}_i$ with a probability that is the target probability $p_i$ normalized by the magnitude of the noise $\mathbf{r}_i$. The solved mechanism is intuitive and interpretable.

\section{Discussion, Limitations, and Future Work}
\label{discussion}


\myparatight{Generalizing AttriGuard to defend against other inference attacks} We believe that there are many opportunities for both the adversarial machine learning community and the security and privacy community to leverage adversarial machine learning to defend against inference attacks in various domains. For the adversarial machine learning community, there are opportunities to develop new adversarial machine learning methods that consider the unique privacy and utility-loss challenges. 
For the privacy community, adversarial machine learning brings new opportunities to achieve better privacy-utility tradeoffs. For the security community, adversarial machine learning brings new opportunities to enhance system security such as designing more secure and usable CAPTCHAs as well as mitigating side-channel attacks. 
Specifically, we envision that AttriGuard's two-phase framework can be applied  to defend against other inference attacks, e.g., the ones we discussed in Section~\ref{sec:attacks}. However, Phase I of AttriGuard should be adapted to different inference attacks, as different inference attacks may have their own unique privacy, security, and utility requirements on the representative noise vectors.  Phase II can be used to satisfy the utility-loss constraints via randomly sampling a representative noise vector according to a certain probability distribution. We note that some recent studies~\cite{Inci18,Imani19} have tried to leverage adversarial examples to defend against website fingerprinting attacks and side-channel attacks. However, they did not consider the utility-loss constraints, which can be satisfied by extending their methods using Phase II of AttriGuard. Moreover, recent studies~\cite{Meng18,Quiring19} have explored adversarial example based defenses against author identification attacks for programs.

\myparatight{Data poisoning attacks based defenses} Other than adversarial examples, we could also leverage data poisoning attacks~\cite{biggio2012poisoning,Jagielski18,poisoningattackRecSys16,YangRecSys17,munoz2017towards,shafahi2018poison,Suciu18,graphrec} to defend against inference attacks. Specifically, an attacker needs to train an ML classifier in inference attacks. For instance, in attribute inference attacks on social networks, an attacker may train a classifier via collecting a training dataset from users who disclose both public data and attribute values. In such scenarios, the defender could inject fake users with carefully crafted public data and attribute values to poison the attacker's training dataset such that the attacker's learnt classifier is inaccurate. In other words, the defender can perform data poisoning attacks to the attacker's classifier. For instance, an online social networking service provider could inject such fake users to defend against inference attacks performed by third-party attackers.

\myparatight{Adaptive inference attacks} We envision that there will be an arms race between attackers and defenders. Specifically, an attacker could adapt its attacks when knowing the defense, while a defender can further adapt its defense based on the adapted attacks.  For instance, an attacker could  first detect the noise added to the public data or detect the fake users, and then the attacker performs inference attacks. Jia and Gong tried  a low-rank approximation based method to detect the noise added by AttriGuard and AttriGuard is still effective against the method. However, this does not mean an attacker cannot perform better attacks via detecting the noise. 
An attacker could also leverage fake-user detection (also known as Sybil detection) methods (e.g.,~\cite{Yu06,sybilrank,Wang13Clickstream,sybilbelief,sybilscar,GANG,sybilfuse,sybilblind,graphsec,mobilesybil}) to detect and remove the fake users when the defender uses data poisoning attacks as defenses.  
We believe it is an interesting future work to systematically study the possibility of detecting noise and fake users both theoretically and empirically. We note that detecting noise is different from \emph{detecting adversarial examples}~\cite{detection1,xu2017feature,MagNet,HeICLR18}, because detecting adversarial examples is to detect whether a given example has attacker-added noise or not.  However, detecting adversarial examples may be able to help perform better  inference attacks. Specifically, if an attacker detects that a public data vector is an adversarial example, the attacker can use a defense-aware  inference attack for the public data vector, otherwise the attacker can use a defense-unaware attack.    

Moreover, an attacker could also use  classifiers, which are more robust to adversarial examples, to perform inference attacks. Jia and Gong evaluated three  robust classifiers: adversarial training~\cite{Goodfellow:ICLR:14b},  defensive distillation~\cite{Papernot16Distillation}, and region-based classification~\cite{region}. They showed that AttriGuard is still effective for attacks using such robust classifiers. 
As the adversarial machine learning community develops more robust classifiers (e.g.,~\cite{Lecuyer19,Cohen19,Wang18}), an attacker could leverage them for inference attacks. However, we speculate that robust classifiers are always vulnerable to adversarial examples that have large enough noise. In other words, we could still leverage adversarial examples to defend against  inference attacks, but we may need larger noise (thus larger utility loss) for the public data when the attacker uses a classifier that is more robust  to adversarial examples. 

\myparatight{Transferability} Transferability of adversarial examples is key to the success of adversarial example based defenses against inference attacks. Therefore, it is important to generate transferable adversarial examples. To enhance transferability, the defender can add larger noise to the adversarial examples (thus larger utility loss) or generate adversarial examples based on an ensemble of classifiers~\cite{liu2016delving}.

\section{Conclusion}
ML-equipped inference attacks pose growing privacy and security threats to users and systems in various domains.  
Attackers rely on the success of ML, but they also share the limitations of ML. In particular, ML classifiers are vulnerable to adversarial examples. 
In this chapter, we discuss the opportunities and challenges of turning the weaknesses of ML into weapons to defend against inference attacks. For instance, we can add carefully crafted noise to the public data to turn them into adversarial examples such that attackers' classifiers make incorrect predictions for the private data. There are many opportunities and challenges for both the adversarial machine learning community and the privacy and security community to study adversarial machine learning based defenses against inference attacks. 
\section*{Acknowledgements} 
This work was supported by NSF grant No. 1801584.


{
\balance{
\bibliographystyle{unsrt}
\bibliography{refs}
}}



%
\end{document}